\documentclass[twocolumn]{aastex631}
\usepackage{verbatim}
\usepackage{amsmath}
\usepackage{graphicx}
\usepackage{subfigure}
\usepackage{bm}
\usepackage{xcolor}
\shorttitle{The Impact of Accretion on FRB Radiation Mechanisms}
\shortauthors{Yao \& Deng}
\graphicspath{{./}{figures/}}
\begin{document}
\title{The Impact of Accretion on FRB Radiation Mechanisms in Binary Systems: Constraints and Implications}
\author{Gong-Yu Yao}

\author[0000-0003-0471-365X]{Can-Min Deng}

\affiliation{
	Guangxi Key Laboratory for Relativistic Astrophysics, Department of Physics, Guangxi University, Nanning 530004, China; dengcm@gxu.edu.cn}

\begin{abstract}
	Fast Radio Bursts (FRBs) are intense, millisecond-duration radio transients that have recently been proposed to arise from coherent radiation mechanisms within the magnetosphere of neutron stars. Observations of repeating FRBs, including periodic activity and large variations in Faraday rotation measures, suggest that these bursts may have binary system origins, with massive companion. In this work, we investigate how accretion from a massive companion influences the FRB radiation within the magnetosphere of the neutron star. Focusing on two widely accepted pulsar-like coherent radiation mechanisms, we establish the parameter space for neutron stars that allows  FRB generation, even in the presence of accreted matter. Our analysis shows that coherent curvature radiation is only viable within a narrow range of parameters, while the magnetic reconnection mechanism operates across a broader range. In both cases, the neutron star must possess a strong magnetic field with strength $\gtrsim 10^{13}$ G. These findings at least indicate that the central engines responsible for producing observable FRBs in binary systems are indeed magnetars.
 
\end{abstract}

\section{Introduction} \label{sec:intro}
Fast radio bursts (FRBs) are highly energetic, millisecond-duration radio transients that have drawn significant attention since their discovery \citep{2021SCPMA..6449501X, 2022A&ARv..30....2P, 2023RvMP...95c5005Z, 2024Ap&SS.369...59L}. Their extremely high brightness temperatures, exceeding $10^{35}$ K \citep{2021Univ....7...56L, 2023RvMP...95c5005Z}, indicate that the radiation mechanisms must be coherent, as incoherent processes are limited by a maximum brightness temperature of around $10^{13}$ K. The discovery of repeating FRBs has confirmed that at least some FRBs originate from non-catastrophic events \citep{Spitler_2016, 2019ApJ...885L..24C, 2019Natur.566..235C, Kumar_2019, Luo_2020, 2020Natur.582..351C, Bhardwaj_2021, Niu_2022}, suggesting the presence of long-lived sources capable of producing multiple bursts over time.

A major breakthrough in understanding FRBs came with the detection of FRB 20200428, which was associated with a Galactic magnetar, SGR 1935+2154 \citep{2020ApJ...898L..29M, 2020Natur.587...54C, 2020Natur.587...59B, 2021NatAs...5..372R, 2021NatAs...5..401T, Li_2021}. This association provides compelling evidence that magnetars—highly magnetized neutron stars—are likely progenitors of at least some FRBs. Following this discovery, numerous theoretical models have been proposed to explain FRB emission from magnetars, {including mechanisms such as coherent curvature radiation \citep{2014PhRvD..89j3009K, 2017MNRAS.468.2726K,Yang_2018,2020ApJ...901L..13Y, 2020MNRAS.494.2385K,Cooper_2021}, coherent inverse Compton scattering \citep{2022ApJ...925...53Z, Qu_2024}, magnetic reconnection \citep{2020ApJ...897....1L,2022ApJ...932L..20M} and synchrotron maser emission \citep{2014MNRAS.442L...9L, 2017ApJ...843L..26B, 10.1093/mnras/stz700}}. These models are broadly classified into two categories based on the location of the radiation: pulsar-like models (close-in models), where the radiation originates within the magnetosphere \citep{2017MNRAS.468.2726K,Yang_2018, Wadiasingh_2019, 2020MNRAS.494.2385K,Wang_2020,2020ApJ...897....1L,Lyutikov_2021, Yang_2021, Cooper_2021,2022ApJ...925...53Z,Qu_2024}, and GRB-like models (far-away models), where the emission occurs in relativistic shocks beyond the magnetosphere \citep{2014MNRAS.442L...9L, 2017ApJ...843L..26B,2018ApJ...868L...4M, 10.1093/mnras/stz700,10.1093/mnras/stz640, 2020ApJ...899L..27M, 2020ApJ...900L..26W, Xiao_2020, 2020ApJ...896..142B}.  
The diverse polarization angle swings observed in some bursts are strong evidence that FRBs may originate from the magnetosphere, as the case of pulsar emission \citep{2020Natur.586..693L,2023ApJ...943...47L,2024arXiv240804401Z,2024ApJ...972L..20N,2024arXiv241114784B}.

Among repeating FRBs, several sources exhibit periodic activity. Notably, FRB 20180916B shows a 16.35-day period \citep{2020Natur.582..351C}, and FRB 20121102 displays a 157-day period \citep{2020MNRAS.495.3551R, 2021MNRAS.500..448C}. These periodicities suggest that some FRBs may originate from binary systems \citep{2020ApJ...893L..26I,2020ApJ...893L..39L,2021ApJ...918L...5L,2021ApJ...922...98D,2022NatCo..13.4382W, 2023ApJ...957....1X, 2023A&A...673A.136R,Lan_2024}, where the observed modulation is caused by orbital motion. Further supporting this hypothesis, FRB 20200120E was localized to a globular cluster in the nearby galaxy M81 \citep{2021ApJ...910L..18B, 2022Natur.602..585K}. The old age of globular clusters implies that the progenitor system likely involves either an old neutron star or a compact binary merger \citep{2020ApJ...890L..24Z, 2021ApJ...917L..11K, 2022MNRAS.510.1867L}.

In addition to periodic activity, several FRBs have shown erratic variations in their Faraday rotation measures (RMs), characterized by large magnitude changes and even sign reversals. For instance, FRB 20190520B \citep{2022NatCo..13.4382W, 2023Sci...380..599A} and FRB 20201124A \citep{2022NatCo..13.4382W} exhibit complex RM variations, which are thought to result from interactions between the FRB source and a dense plasma environment. \citet{2022NatCo..13.4382W} proposed that FRB 20201124A and FRB 20190520B may originate from magnetar/Be star binary systems. Furthermore, \citet{2023ApJ...942..102Z} and \citet{2023A&A...673A.136R} suggested that the complex RM behavior observed in some repeating FRBs could be explained by binary systems, where interactions with the companion’s stellar wind or decretion disk influence the surrounding plasma.

While most existing FRB models focus on isolated neutron stars  \citep{2020MNRAS.494.2385K,2020ApJ...897....1L,2022ApJ...925...53Z}, the presence of a binary companion introduces new complexities, particularly in terms of accretion processes \citep{1952MNRAS.112..195B, 1973ApJ...179..585D,1989PASJ...41....1N}. In a binary system, matter from the companion star can accrete onto the neutron stars, altering the plasma environment in the magnetosphere \citep{1989Natur.342..656S,1997MNRAS.284..311K,2001ApJ...557..958C}. This accreted matter can impact the radiation mechanisms responsible for generating FRBs, potentially suppressing coherent emission or changing the conditions required for such processes to occur. Therefore, the effects of accreted matter on the magnetosphere and the resulting implications for FRB radiation mechanisms must be carefully considered.

  In this study, we investigate the impact of accretion in binary systems on close-in radiation mechanisms, with a particular focus on coherent curvature radiation \citep{2020MNRAS.494.2385K} and magnetic reconnection \citep{2020ApJ...897....1L}. These mechanisms are highly sensitive to the plasma environment within the magnetosphere and may be significantly affected by the presence of accreted material. We examine the conditions under which these mechanisms remain viable in the presence of accretion and explore the parameter space necessary for neutron stars in binary systems to produce observable fast radio bursts (FRBs). By considering different evolutionary stages of the neutron star, we aim to establish a comprehensive framework for understanding the feasibility of FRB radiation mechanisms in binary systems with accreting neutron stars.  { Our results indicate that, in general, only in the ejector stage does the neutron star's magnetic field possess sufficient strength to expel material, allowing the coherent curvature radiation mechanism to remain viable as the plasma density remains low. In contrast, the magnetic reconnection model operates over a broader parameter space, extending into the propeller stage, where material is partially expelled due to centrifugal forces. This is reflected in the larger viability region for magnetic reconnection compared to coherent curvature radiation, which is confined to a narrow parameter space within the ejector phase.}

\section{Accretion onto neutron star}\label{sec:acc}
   {In this section,  we discuss the specific conditions under which a neutron star in the binary system accretes material from its massive companion reaching its surface. We consider the main sequence star as companion to the neutron star, and the density of the matter accreted onto the neutron star in the FRB emitting region is the focus of our attention.
In general, the accretion evolution of the neutron star in binary system is determined by both its own properties and that of the companion. If the companion is a high-mass star, the mass transfermation  would be mainly achieved through the process of wind captured  \citep{1952MNRAS.112..195B, 1973ApJ...179..585D,1989PASJ...41....1N,2012MNRAS.420..216S}, because  the main sequence companion never fills its Roche lobe during the evolution of the binary \citep{2024Univ...10..205A}.}

\subsection{Mass transfer rate $\dot{M}_{\rm T}$}
{For a high-mass O/B-type main sequence star with mass of $M_{\rm c}\gtrsim 10M_{\odot}$, the orbital semi-major axis is estimated as}
\begin{eqnarray}
a & \simeq & \left[\frac{GM_{\rm c}}{\left(2\pi/P_{\rm orb}\right)^{2}}\right]^{1/3}\nonumber\\
&=& 2.9 \times 10^{12} ~M^{1/3}_{\rm c,1}P_{\rm orb,1}^{2/3}\,{\rm cm},
\end{eqnarray}
where $M_{{\rm c},x}=10^{x}M_{\odot}$, $P_{{\rm orb},x}=10^{x}$ day. It is much larger than the gravitational capture radius $ R_{\rm G}$ (see equation (\ref{RG})). Therefore, the density of the stellar wind at the capture radius of the neutron star is $\rho_{\rm w} (a) = {\dot{M}_{\rm loss}}/{4\pi a^{2}v_{\rm w}}$, where $\dot{M}_{\rm loss}$ is the mass loss rate of the companion star and $v_{\rm w}$ is the velocity of the wind at infinity.  For a typical massive main sequence  star,  the mass loss rate $\dot{M}_{\rm loss} \sim  10^{-8}\ M_{\odot}\  {\rm yr}^{-1}$ \citep{1981ApJ...251..139S, 2014A&A...564A..70K} and the wind velocity  $v_{\rm w}\sim 10^{8} ~{\rm cm}\,{\rm s}^{-1}$ \citep{1995ApJ...455..269L} are usually expected \citep{2014ARA&A..52..487S,2024Univ...10..205A}. 
Then the rate of the neutron star capture the stellar wind from the massive companion can be estimated as  \citep{2002apa..book.....F}
\begin{eqnarray}\label{dotMB}
\dot{M}_{\rm T} &=& 4\pi R_{\rm G}^{2}\rho_{\rm w}v_{\rm w}\nonumber\\
&\simeq& 1.6 \times 10^{-12}\dot{M}_{\rm loss,-8}\nonumber\\
&\times&M^{-2/3}_{\rm c,1} P_{\rm orb, 1}^{-4/3}v_{\rm w,8}^{-4}\,M_{\odot}\,{\rm yr}^{-1},
\end{eqnarray} 
{where $\dot{M}_{\rm loss,-8}=10^{-8} M_{\odot}\,{\rm yr}^{-1}$, $v_{\rm w,8}=10^{8} ~{\rm cm}\,{\rm s}^{-1}$.
One sees that a tiny fraction of the wind would be captured by the neutron star.}

\subsection{The characteristic radii}\label{radii}
 {According to different conditions, the accretion state of the neutron star can be divided into the following evolutionary stages: ejector, propeller, accretor, and georotator \citep{1992ans..book.....L, 2024Galax..12....7A}. The amount of material captured by the neutron star from its companion that can reach the neutron star's surface, i.e., the accretion rate $\dot{M}_{\rm X}$, can vary significantly at different evolutionary stages.
	The transition of evolutionary stages can often be specified in terms of the equality of some characteristic radii. }
They are the light cylinder radius $R_{\rm L}$, gravitational capture radius $R_{\rm G}$, corotation radius $R_{\rm co}$, magnetospheric radius $R_{\rm m}$, and Shvartsman radius $R_{\rm Sh}$.
\par
    Magnetic field lines corotate with the neutron star. Thus, the linear velocity of a field line grows (in the equatorial plane) as $\Omega R$, where $R$ is the distance from the center of neutron star and $\Omega = 2\pi/P$ is spin frequency, $P$ is the spin period. As this velocity is limited by the speed of light $c$, there is a critical distance called the light cylinder radius,
\begin{eqnarray}\label{RL}
    R_{\rm L}=\frac{c}{\Omega}\simeq 4.7\times 10^{9}P_{0}\, {\rm cm}.
\end{eqnarray}
\par
    The equality of kinetic energy and the absolute value of potential energy of the matter surrounding a compact object defines the gravitational capture radius (i.e. the Bondi radius),
\begin{eqnarray}
    R_{\rm G}=\frac{2GM_{*}}{v_{\rm{w}}^2}\simeq 3.7\times 10^{10}v_{\rm{w,8}}^{-2}\, {\rm cm}\label{RG},
\end{eqnarray}
    here $G$ is the Newton constant, $M_{*}=1.4 M_{\odot}$ is the neutron star mass, and $v$ is the velocity relative to the medium.
\par
    Plasma frozen in the magnetosphere corotates with the maximum velocity $\Omega R$. If this value is larger than the local Keplerian velocity $\sqrt{GM_{*}/R}$, then a centrifugal barrier prevents accretion down to the neutron star surface. Equality of the linear and Keplerian velocity defines the corotation radius,
\begin{eqnarray}
    R_{\rm co}=\left(\frac{GM_{*}}{\Omega^{2}}\right)^{1/3}\simeq 1.6 \times 10^{8}P^{2/3}_{0}\, {\rm cm}\label{Rco}.
\end{eqnarray}
\par
    The gravitationally captured matter will not fall directly onto the neutron star surface, at magnetospheric radius $R_{\rm m}$ the matter can be stopped by the magnetic field. The magnetospheric radius $R_{\rm m}$ might be calculated differently in the case of disk and spherical accretion. In addition, the radius is calculated differently depending on the relative value of $R_{\rm m}$ and $R_{\rm G}$. {For spherical accretion flow with a rate $\dot{M_T}$ }and $R_{\rm m} < R_{\rm G}$, we have
\begin{eqnarray}\label{Rm}
    R_{\rm m}&=&\left(\frac{\mu^{2}}{2\dot{M_T}\sqrt{2GM_{*}}}\right)^{2/7}\nonumber\\
    &\simeq& 2.6\times 10^{10}B_{14}^{4/7} M^{4/21}_{\rm c,1} P_{\rm orb, 1}^{ 8/21} \dot{M}_{\rm loss,-8}^{-2/7}v_{\rm w,8}^{8/7} \, {\rm cm}.
\end{eqnarray}
    This radius is also called the magnetospheric radius, $R_{\rm A}$. Where $\mu=BR^{3}_{*}$ is a magnetic moment, $B$ is the equatorial surface dipolar field, $R_{*}\sim 10^{6}\,{\rm cm}$ is the radius of neutron star. {Note, that $ \dot{M}$ appears in the equation as the mass rate at which the neutron star captures material from its companion. }
\par
    Shvartsman radius $R_{\rm Sh}$ is determined by the balance between the pressure of relativistic particles wind and external medium. For $R_{\rm Sh} > R_{\rm G}$ and standard magneto-dipole rate
    of losses, we have
\begin{eqnarray}\label{RSh}
    R_{\rm Sh}&=&\left(\frac{2\mu^{2}(GM_{*})^{2}\Omega^{4}}{3\dot{M_T}v^{5}_{\rm{w}}c^{4}}\right)^{1/2}\nonumber\\
    &\simeq& 6.9 \times 10^{11}B_{14}P_{0}^{-2} \dot{M}_{\rm loss,-8}^{-1/2}
    M^{1/3}_{\rm c,1} P_{\rm orb, 1}^{2/3}v_{\rm{w,8}}^{-1/2}\, {\rm cm}~.
\end{eqnarray}

\subsection{Evolutionary stages and the number density of accreted matter in the magnetosphere}
\subsubsection{Ejector stage}
    At the ejector stage, the external {medium ( the matter from companion)} is stopped at the Shvartsman radius $R_{\rm Sh}$, which is greater than both the light cylinder radius $R_{\rm L}$ and the gravitational capture radius $R_{\rm G}$. {According to equation (\ref{RL}) (\ref{RG})  (\ref{RSh}), one can defined an equilibrium spin period
     $P_{\rm EP}$ for $R_{\rm Sh}=$ max($R_{\rm G}$, $R_{\rm L}$),}
    \begin{eqnarray}\label{P_EP}
    P_{\rm EP} 
    &\simeq&
    \left\{
    \begin{aligned}
    4.3 &B_{14}^{1/2}  \dot{M}_{\rm loss,-8}^{-1/4}M_{\rm c,1}^{1/6}\\
    &\times P_{\rm orb, 1}^{1/3}v_{\rm w,8}^{3/4}\,{\rm s}, \quad  {R_{\rm G}>R_{\rm L}}\\
    5.3 &B_{14}^{1/3}  \dot{M}_{\rm loss,-8}^{-1/6}M_{\rm c,1}^{1/9}\\
    &\times P_{\rm orb, 1}^{2/9}v_{\rm w,8}^{-1/6}\,{\rm s}, \quad {R_{\rm G} \leq R_{\rm L}}
    \end{aligned}
    \right.
    \end{eqnarray}
    In this stage, one has $P<P_{\rm EP} $, the neutron star can be considered isolated and{ the accretion rate onto the neutron star is $\dot{M}_{\rm X}=0$. }
\par
\subsubsection{Propeller stage}
    When the neutron star spin down to $P>P_{\rm EP}$,  the $R_{\rm m}$ would  decrease to be less than  $R_{\rm L}$, such that the external matter could penetrate the light cylinder.  If $R_{\rm m} > R_{\rm co}$ is also satisfied, the matter would ultimately be thrown outward due to the centrifugal force acting against the gravity, and therefore the matter could not freely reach the surface of the neutron star. This corresponds to the propeller stage. 
    However, this does not mean that there is no material that can be accreted onto the neutron star surface at all. As pointed out by \cite{1999ApJ...520..276M}, the centrifugal acceleration at some polar angle $\theta$ from the spin axis is able to be smaller than gravitational acceleration, which the matter blow a critical angle $\theta_{\rm c}$ can be accreted onto the neutron star. The maximum polar angle below which the gravitational acceleration ($=R_{\rm m}\Omega^{2}_{\rm K}(R_{\rm m})$) wins over the centrifugal force is given by $ \sin{\theta_{\rm c}}={\Omega_{\rm K}(R_{\rm m})}/{\Omega}$,
    where the $\Omega_{\rm K}(R_{\rm m})$ is the Kepler velocity in the magnetospheric radius. For $\theta_{\rm c}\ll 1$ this simplifies to $\theta_{\rm c}\simeq {\Omega_{\rm K}(R_{\rm m})}/{\Omega}$. 
    Only the matter accreted by the neutron star between $\theta=0$ and $\theta_{\rm c}$ can overcome the centrifugal barrier and reach the surface of the neutron star. The fraction of mass inflow that can accrete onto the neutron star is then given by $f={\dot{M}_{\mathrm{X}}}/{\dot{M}_T}=3 \theta_{\rm c}^{4}/8$,
    where $\rho(R,\theta)\simeq \rho(R)$ is the density of accreted matter and $v_{\rm r}(R,\theta)\simeq v_{\rm r}(R)\sin ^{2} \theta$ is the radial infall velocity at angle $\theta$.   Thus, the density of the accreted material in the magnetosphere may be calculated as \citep{1999ApJ...520..276M}, which evolves with radius.
   
    \begin{eqnarray}
    n_{\rm P} &=& \frac{\dot{M}_{\rm X}}{2 m_{\rm p}\int_{0}^{\theta_c} 2 \pi R^{2}  v_{\rm r}(R) \sin ^{3} \theta d \theta}\nonumber\\
    &\simeq& 3.4 \times 10^{11}\dot{M}_{\rm loss,-8}M_{\rm c,1}^{-2/3}P_{\rm orb, 1}^{-4/3}\nonumber\\
    &\times& v_{\rm w,8}^{-4}R_{8}^{-3/2}\,{\rm cm}^{-3},
    \end{eqnarray}
    where $m_{\rm p}$ is the mass of proton as the accreted matter, $v_{\rm r}(R)$ is the drop velocity of the accreted matter, generally taken as free-fall velocity, $v_{\rm r}(R)\approx v_{\rm ff}= \sqrt{2GM_{*}/R}$. And $\theta$ is the angle that accreted matter can fall onto the neutron star surface, here we just consider the case where the magnetic axis and rotation axis coincide.
   It is important to note that when the neutron star continues to spin down until $R_{\rm m} < R_{\rm co}$ the accretion of the neutron star will enter the next phase.  Again, one can defined an equilibrium spin period
   $P_{\rm PA}$ for $R_{\rm m}=R_{\rm co}$,
   \begin{eqnarray}\label{P_PA}
   P_{\rm PA}&=& \frac{2^{5/14}\pi\mu^{6/7}}{\dot{M_T}^{3/7}(GM_{*})^{5/7}}\nonumber\\
   &\simeq&
   1.9 \times 10^{3}B_{14}^{6/7}\dot{M}_{\rm loss,-8}^{-3/7}M_{\rm c,1}^{2/7}
    P_{\rm orb, 1}^{4/7}v_{\rm w,8}^{12/7}\,{\rm s}.
   \end{eqnarray}
   In other words, when $P>P_{\rm PA}$, the neutron star will leave the propeller stage and enter the accretor stage.
\par
\subsubsection{Accretor stage}
     {In this stage, the captured matter can be freely accreted onto the surface of neutron star. Therefore, one has $\dot{M}_{X}\simeq \dot{M}_T$. The number density of the accreted  matter in the magnetosphere  may be calculated as \citep{2002apa..book.....F}}
    \begin{eqnarray}
    n_{\rm A} &=& \frac{\dot{M}_{\rm X}}{2 m_{\rm p}\int_{0}^{\beta} 2 \pi R^{2}  v_{\rm r}(R) \sin ^{3} \theta d \theta}\nonumber\\
    &\simeq&
    6.2 \times 10^{20}B_{14}^{8/7}\dot{M}_{\rm loss,-8}^{3/7} M_{\rm c,1}^{-2/7}\nonumber\\
   &\times& P_{\rm orb, 1}^{-4/7}
    v_{\rm w,8}^{-12/7}R_{8}^{-3/2}\,{\rm cm}^{-3},
    \end{eqnarray}
    where $\beta$ is the angle of accreting polecaps given by $\beta \simeq (R_{*}/R_{\rm m})^{1/2}$ \citep{2002apa..book.....F}. It should be noted that, at this stage, it is also necessary to keep $R_{\rm m} < R_{G}$ in order for matter to be accreted freely onto the neutron star, in addition to $R_{\rm m} < R_{\rm co}$.
    Otherwise, a magnetosphere larger than the capture radius would significantly inhibit the accretion of the captured matter, which corresponds to the georotator stage \citep{1975A&A....39..185I,1992ans..book.....L}.
    
\par
\subsubsection{Georotator stage}
    Georotator is related to the fact that the shape of the magnetosphere and some processes are similar to the case of the Earth magnetosphere in the Solar wind. Such a situation can be realized if the velocity of matter relative to the neutron star is too high, if the magnetic field is too large, or if matter density is too low.  In this case, the matter flows around the magnetosphere as the gravitational influence of the neutron star is too small at $R > R_{\rm m}$ due to $R_{\rm m} > R_{G}$, { i.e., $B>B_{\rm G}$,
\begin{eqnarray}\label{B_G}
    B_{\rm G} &=& \left(\frac{2GM_{*}}{v_{\rm w}^{2}}\right)^{7/4}\left(2\dot{M}_{\rm T}\sqrt{2GM_{*}}\right)^{1/2}\frac{1}{R_{*}^3}\nonumber\\
    &\sim& 1.9\times 10^{14} \dot{M}_{\rm loss,-8}^{1/2}M^{-1/3}_{\rm c,1} P_{\rm orb, 1}^{-2/3}v_{\rm w,8}^{-11/2}\,{\rm G},
\end{eqnarray}}
    where $R_{\rm m} = R_{*} ({B^{2}}/{4\pi\rho_{\rm w}v_{\rm w}^{2}})^{1/6}$ is re-defined by the balance of the magnetic pressure of the magnetosphere against the ram pressure of the incoming stellar wind. However, \cite{2001ApJ...561..964T} found that a small fraction of the stellar wind would still be captured by interacting with the magnetosphere. 
    The capture rate of the wind by  the neutron star is calculated by $\dot{M}_{\rm T}=4\pi R_{\rm m}^{2}\rho_{\rm w}v_{\rm w}$, and a fraction of $f \sim 10^{-3}$ could finally be accreted  onto the neutron star surface \citep{2001ApJ...561..964T}.  Therefore, ,in this case, the number density of the accreted  matter in the magnetosphere  is 
  \begin{eqnarray}\label{n_G}
  n_{\rm G} &=& \frac{f \dot{M}_{\rm{T}}B^{2/3}}{\pi m_{p}(2GM_{*})R^{3/2}(4\pi \rho_{\rm w}v_{\rm w}^{2})^{1/3}}\nonumber\\
  &\simeq& 8.8\times 10^{17}f_{-3}B^{4/3}_{14}\dot{M}_{\rm loss,-8}^{1/3}\nonumber\\
  &\times&M_{\rm c,1}^{-2/9}P_{\rm orb, 1}^{-4/9}v_{\rm w,8}^{-2/3}R_{8}^{-3/2}~{\rm cm}^{-3}.
\end{eqnarray}
As one can see, the accretion onto the neutron star in the georotator stage is weaker than that in the accretor stage, but still much stronger than that in the propeller stage.

\section{Limitations on neutron star's parameters that can produce observable FRBs }\label{FRB Coherent Emission from decay of Alfvén Waves}
It can be seen from the previous discussion that except in the ejector stage, neutron stars in binary systems will accret materials from their companion stars, thus polluting their own magnetosphere. This is bound to cause potential interference with the generation of FRBs in the magnetosphere. Providing that diverse polarization angle swings observed in some FRBs \citep{2020Natur.586..693L,2024ApJ...972L..20N,2024arXiv241114784B}, it is argued that they are  likely generated by coherent  radiation within the magnetosphere \citep{2021Univ....7...56L, 2023RvMP...95c5005Z}, although little is known about the mechanism that provides the bunched electrons and conditions involved for producing the coherent radiation. Among the radiation mechanisms proposed in the literature, the one put forward by \cite{2020MNRAS.494.2385K} is perhaps the most widely discussed for FRB generation within the magnetosphere. Alternatively, magnetic reconnection occurring at the light cylinder has also been proposed as a potential mechanism for FRB production \citep{2020ApJ...897....1L}. Below, we will mainly conduct detailed discussions based on these two radiation models, respectively.

\subsection{Decayed Alfvén Waves model}
In this scenario, the Alfvén wave may be due to the sudden crustal motion or the starquake of the neutron star is emitted from the surface of the neutron star and propagates outward along the magnetic field lines at the polar cap, which has a strong magnetic field amplitude  \citep{2020MNRAS.494.2385K}.  Alfvén waves have non-zero current along the magnetic field lines unless the wave vector is perfectly parallel to the magnetic field lines. Electric current is carried by electrons and positrons moving in opposite directions. This countercurrent motion of $e^{\pm}$ is affected by the two-flow instability, which leads to the formation of a large number of charged particle bunches moving along the magnetic field lines. In an isolated neutron star, where the plasma density decreases with distance from the neutron star surface, the velocity of the particles increases to carry the current density required for the Alfvén wave. When the plasma density at a certain distance in the neutron star magnetosphere is below a critical value, the plasma cannot support the current required by the Alfvén wave even if $e^{\pm}$ moves at the speed of light. A strong electric field is then generated, and the displacement current associated with this field compensates for the lack of plasma current density. Like the plasma current, the electric field has a component along the static magnetic field. This electric field forces the electrons and positrons in the clumps to move along the magnetic field lines in opposite directions with a high Lorentz factor, then these charged particle bunches radiate coherent curvature as they move along the magnetic field lines to emit FRBs.
\par
This coherent curvature radiation mechanism of Alfvén wave decay requires the Alfvén wave packet to reach the charge starved region to generate the strong parallel electric field to accelerate the particle bunches. However, the particle number density in the open filed lines region will be significantly increased by the accreted matter if the neutron star have a massive companion, which may have an effect on the propagation of the Alfvén wave as well as on the radiation mechanism. 
\par
An Alfvén wave-packet launched at the surface of a neutron star with amplitude $B_{\rm A}\sim\xi B$ and luminosity $L_{\rm aw}\simeq 10^{42}(\xi_{-4} B_{14})^{2}R_{*}^{2}c\,{\rm erg}\,{\rm s}^{-1}.$, $\xi\leq 10^{-4}$ is for the wave to become nonlinear in the region of $R<10^{9}\,{\rm cm}$ \citep{2020MNRAS.494.2385K,2022MNRAS.515.2020Q}. Therefore, the critical plasma density in the charge starved region is \citep{2020MNRAS.494.2385K}
\begin{eqnarray}\label{eq:ncR}
n_{\rm c}(R) &=& \frac{k_{\rm aw\perp}B_{\rm A}}{8\pi q}=\frac{L_{\rm aw, 42}^{1/2}R_{*}^{2}}{4q\lambda_{\rm aw\perp}c^{1/2}R^{3}}\nonumber\\
&\simeq& 5.2\times 10^{8}\xi_{-4} B_{14}\lambda_{\rm aw\perp, 4}^{-1}R^{-3}_{8}\,{\rm cm}^{-3},
\end{eqnarray}
where $k_{\rm aw\perp}$ is the wave vector component of the Alfvén wave perpendicular to the background magnetic field $B$. When the Alfvén wave propagates into a region where the plasma density is below this critical density, the Alfvén wave will enter the charge starvation region, where a strong electric field with a non-zero component is formed along the background magnetic field and accelerates the particle clumps produced by the two-stream instability through the coherent curvature radiation to generate the FRB.
\par
All of the above are considered in isolated neutron star, but for neutron stars in binary system, it is likely that there is an accretion process, and it should be considered whether the accretion process will have an effect on the whole radiation mechanism. For this coherent curvature radiation mechanism of Alfvén wave decay, the most important effect comes from the effect of accretion on the outward propagating Alfvén wave, i.e., whether the Alfvén wave can reach the charge-starved state in the presence of the accretion. For the accreting neutron star, the total number density of the plasma in the magnetosphere is 
\begin{eqnarray}
n_{\rm tot} 
&=&
\left\{
\begin{aligned}
&n_{\rm A}+n_0\quad \text{for accretor stage},\\
&n_{\rm P}+n_0\quad \text{for propeller stage},\\
&n_{\rm G}+n_0\quad \text{for georotator stage}.
\end{aligned}
\right.
\end{eqnarray}
where $n_0$ is the particle density in neutron star magnetosphere controlled by the magnetic field and the rotation of the neutron star \citep{1969ApJ...157..869G}, and has a minimum number density $n_{\rm GJ}$ (Goldreich Julian density), and the actual particle density is considered to be a multiple of the GJ density, 
\begin{eqnarray}\label{n}
n_0=\mathcal{M}n_{\rm GJ}=\frac{\mathcal{M}\textbf{B}\cdot\boldsymbol{\Omega}}{2\pi qc}\approx\frac{\mathcal{M}B_{14}\Omega }{2\pi qc}\left(\frac{R_{*}}{R}\right)^{3}\nonumber\\
\simeq 7.0\times 10^{6}\mathcal{M}_{1}B_{14}P_{1}^{-1}R^{-3}_{8}\,{\rm cm}^{-3},
\end{eqnarray}
where $\mathcal{M}$ is the multiplicity coefficient \citep{1983AIPC..101..163A}, this value is highly uncertain and is estimated to be anywhere between a few and $\sim 10^{6}$ for neutron star, we set $\mathcal{M}\sim 10$ at the FRB emission radius. 
    
\par
    By combining  the number density at each  stage we have got in the previous section, we found the density at the georotator stage $n_{\rm G}$ and at the accretor stage $n_{\rm A}$ both exceed the critical density $n_{\rm c}$. The density of the propeller stage $n_{\rm P}$ would not be less than the critical density $n_{\rm c}$, unless the following condition was satisfied,
    \begin{figure*}
    	\centering
    	\subfigure{\includegraphics[width=0.8\linewidth]{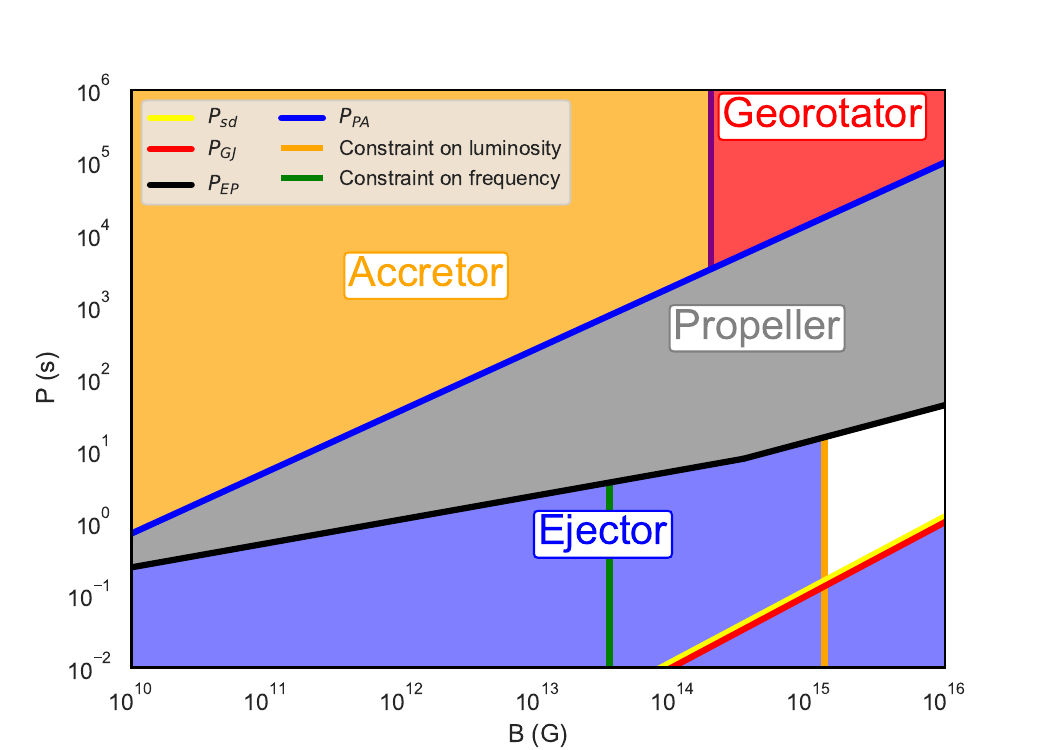}\label{fig:es2}}
    	\caption{The neutron star parameters for the formation of charge-starved region and the generation of ${\rm GHz}$ and characteristic luminosity FRBs. Here, the mass of companion star $M_{\rm c}\sim 10M_{\odot}$, the mass loss rate of companion star $M_{\rm loss}\sim 10^{-8} M_{\odot}\,{\rm yr}^{-1}$, and
    	the period of orbital $\sim 10\,{\rm d}$ are adopted for demonstration. The the blue line is $P_{\rm PA}$ \eqref{P_PA}, above this line is the accretor stage (orange area). In the accretor stage, when the magnetic field is large enough ({purple line \eqref{B_G}}), the accretor stage will be transformed into georotator stage (red area). The black line is $P_{\rm EP}$ \eqref{P_EP}, between blue line and black line is propeller stage (gray area). Under the black line is ejector stage (blue area). In the ejector stage, green line is the constraint on frequency \eqref{BGHz}, the area right of this line is the neutron star parameters that ${\rm GHz}$ FRB can be generated. The orange line is constraint on luminosity \eqref{Blum}, the area right of this line is the magnetar parameters that FRB with characteristic luminosity $(10^{42}\,{\rm erg}\,{\rm s}^{-1})$ can be generated. The red line is the condition that charg-starved region can be formed in the ejector stage \eqref{P_GJ}. The yellow line is the constraint on spin-down $(\tau>1\,{\rm yr})$ of neutron star due to magnetic dipole radiation \eqref{P_sd}. The the white area is the neutron star parameters for the formation of charge-starved region and the generation of ${\rm GHz}$ and characteristic luminosity FRBs.}
    	\label{condition of charge starved}
    \end{figure*}
\begin{eqnarray}\label{L_aw_need}
    L_{\rm aw}&>&\left(\frac{4n_{\rm P}q\lambda_{\rm aw\perp}c^{1/2}R^{3}}{R_{*}^{2}}\right)^{2}\nonumber\\
    &\sim& 1.3\times 10^{48}\dot{M}_{\rm loss,-8}^{2}M_{\rm c,1}^{-4/3}P_{\rm orb, 1}^{-8/3}\nonumber\\
    &\times& v_{\rm w,8}^{-8}R_{8}^{3}\lambda_{\rm aw\perp,4}^{2}\,{\rm erg}\,{\rm s}^{-1}.
\end{eqnarray}
    Such a high luminosity of the Alfvén wave means that the corresponding FRB  luminosity reaches $> 10^{45}\,{\rm erg}\,{\rm s}^{-1}$ (see equation \ref{L_FRB} ), which is well larger than the characteristic  luminosity  of the observed FRBs.
    Therefore, for the propeller stage, even if it is possible to make the density in this stage less than the critical density, it is not possible to produce a suitable FRB that have been observed.
\par
   In the ejector stage, no material could be accreted into the magnetosphere of neutron star, the density in the emission region only is $n$.  Here, the rotation period of the neutron star should satisfies $P<P_{\rm EP}$ (see equation \eqref{P_EP}).  Meanwhile, the generated FRBs should be in the ${\rm GHz}$ range. The Lorentz factor of the radiating particle clump is
     \citep{2020MNRAS.494.2385K}
\begin{eqnarray}
    \gamma\sim\frac{R_{\rm B}}{R}\left[\frac{3E_{\parallel}}{2qn_{\rm c}l_{\parallel}}\right]^{1/2}\label{gammap},
\end{eqnarray}
    where the $R_{\rm B}\sim 0.8R/\theta_{B}$ is the curvature radius of a dipole magnetic field line at the FRB emission radicus $R$, $\theta_{B}<1$ is the polar angle that is measured wrt to the magnetic axis \citep{2020MNRAS.494.2385K,2022MNRAS.515.2020Q}. The electric field $E_{\parallel}$ is associated with the displacement current because of charge starved that is paralleled the background magnetic field
\begin{eqnarray}
    E_{\parallel} &=& \frac{8\pi qcn_{\rm c}}{\omega_{\rm aw}}=\frac{k_{\rm aw\perp}}{k_{\rm aw\parallel}}B_{\rm A}\nonumber\\
    &\sim& 10^{5} \xi_{-4} B_{14}R^{-3/2}_{8}\, {\rm esu},
\end{eqnarray}
    where the $k_{\rm aw\perp}/k_{\rm aw\parallel} \sim 0.01$ at the FRB emission radius $R\sim 10^{8}\,{\rm cm}$ \citep{2020MNRAS.494.2385K}. And the $l_{\parallel}$ is the longitudinal size of a typical clump
\begin{eqnarray}
    l_{\parallel}&\sim& c\left(\frac{\pi m_{e}}{4q^{2}n_{c}}\right)^{1/2}\nonumber\\
    &\sim& 7.3\times 10^{1} \xi_{-4}^{-1/2} B_{14}^{-1/2}\lambda_{\rm aw\perp, 4}^{1/2}R^{3/2}_{8}\,{\rm cm}.
\end{eqnarray}
    So the equation \eqref{gammap} is rewritten as
\begin{eqnarray}
    \gamma&\sim&\frac{0.8}{\theta_{\rm B}}\left(\frac{3k_{\rm aw\perp}L_{\rm aw, 42}^{1/2}R_{*}^{1/2}}{(\pi c^{3}m_{\rm e}n_{\rm c})^{1/2}k_{\rm aw\parallel}R^{3/2}}\right)^{1/2}\nonumber\\
    &\sim& 7.2\times 10^{2}\theta_{\rm B, -1}^{-1}\xi_{-4}^{1/4} B_{14}^{1/4}\lambda_{\rm aw\perp,4}^{1/4}.
\end{eqnarray}    
    The change of the Lorentz factor leads to the change of frequency of the FRB produced by the coherent curvature radiation mechanism
\begin{eqnarray}
    \nu &=& \frac{c\gamma^{3}}{2\pi R_{\rm B}}\nonumber\\
    &\sim& 2.3\times 10^{9}\theta_{\rm B, -1}^{-2}\xi_{-4}^{3/4} B_{14}^{3/4}\lambda_{\rm aw\perp,4}^{3/4}R_{8}^{-1}\,{\rm Hz}.
\end{eqnarray}
    Therefore, based on the typical radiation frequency of FRBs, magnetic field of the neutron star must be satisfied
\begin{eqnarray}\label{BGHz}
    B&>&3.4\times 10^{13}\theta_{\rm B, -1}^{8/3}\xi_{-4}^{-1}\lambda_{\rm aw\perp,4}^{-1} \nu_9 ^{4/3} R_{8}^{4/3}\,{\rm G}.
\end{eqnarray}
It should be noted that the  ">" sign in the above equation arises from the condition that $\xi < 10^{-4}$.
\par
 Moreover, based on this model, the isotropic  luminosity of the FRBs is related to the magnetic field by 
\begin{eqnarray}\label{L_FRB}
    L_{\rm FRB} &\approx& \frac{16(2 \pi)^{2 / 3} q^{2} c^{1 / 3} R^{5} n_{\rm c}^{2} \ell_{\parallel} \nu^{2 / 3}}{3 R_{B}^{4 / 3}}\nonumber\\
    &\simeq& 5.9\times 10^{39}\xi_{-4}^{2}B_{14}^{2}\lambda_{\rm aw\perp,4}^{-1}R_{8}^{-3/2}\,{\rm erg}\,{\rm s}^{-1}.
\end{eqnarray}
   Thus, we have another lower bound for the magnetic field $B$ for the typical luminosity of a FRB source, $L_{\rm FRB}=10^{42}\,{\rm erg}\,{\rm s}^{-1}$ for example,
\begin{eqnarray}\label{Blum}
    B>1.3\times 10^{15}\xi_{-4}^{-1}L_{\rm FRB,42}^{1/2}\lambda_{\rm aw\perp,4}^{1/2}R_{8}^{3/4}\,{\rm G}.
\end{eqnarray}
    On the other hand, the condition that charge-starved region can be existed in ejector stage is $n_{\rm c}>n$, or that is 
\begin{eqnarray}\label{P_GJ}
    P&>&P_{\rm GJ} = \frac{4\mathcal{M}B_{14}\lambda_{\rm aw\perp}R_{*}}{L_{\rm aw}^{1/2}c^{1/2}}\nonumber\\
    &\simeq& 1.0\times 10^{-2} \mathcal{M}_{1}L_{\rm FRB,42}^{-1/2}B_{14}\lambda_{\rm aw\perp,4}^{1/2}R_{8}^{-3/4}\,{\rm s}.
\end{eqnarray}

 In addition, the neutron star will spin down due to its own magnetic dipole radiation. The  spin down timescale can be calculated as
$\tau =3 c^{3} I P^{2}/(4 \pi^{2} B^{2} R_{*}^{6})$, 
where $I\sim 10^{45}\,{\rm g}\,{\rm cm}^{2}$ is moment of inertia of neutron stars. For FRBs active for more than one year, the rotation period of the neutron star  must be satisfied
\begin{eqnarray}\label{P_sd}
P&>&P_{\rm sd}=\frac{2 \pi B R_{*}^{3}\tau^{1/2}}{3^{1/2} c^{3/2} I^{1/2} }\nonumber\\
&\simeq&1.2\times 10^{-2} B_{14}  \tau_{\rm 1yr}^{1/2} I_{45}^{-1/2}~{\rm s}.
\end{eqnarray}

In summary, the constraints on the parameters of neutron stars are shown in the figure \ref{condition of charge starved}. 
The white region depicted in the figure represents the viable range of neutron star parameters, specifically the magnetic field and rotation period. Notably, it is only a small fraction within the ejector stage, and more precisely, the region within this ejector stage where the magnetic field  is relatively large that can meet these constraints. Within this particular region, the neutron star is capable of forming a charge starved region within the binary system context. It is precisely within this charge starved region that FRBs with characteristic luminosity in the GHz frequency band might potentially be generated. This emphasizes the highly specific and restricted conditions required for the FRB radiation mechanism to operate, highlighting the need for a more detailed understanding of the neutron stars behavior and parameter space in the binary system to fully explain FRB phenomena.

{ In Figure \ref{condition of charge starved}, we set the mass-loss rate $(\dot{M}_{\rm loss})$ and the companion mass $(M_{\rm c})$ to typical values. Note that, both the mass-loss rate and companion mass influence the accretion process by directly altering the mass transfer rate $\dot{M}_{\rm T}\propto \dot{M}_{\rm loss}M_{\rm c}^{-2/3}$, which in turn affects the positions of the purple line \eqref{B_G}, the blue line \eqref{P_PA}, and the black lines \eqref{P_EP} in Figure \ref{condition of charge starved}, as well as the density of accreted material at each stage. However, this effect is not strong enough to alter our conclusions. For massive stars in the main-sequence stage, the mass-loss rate has been estimated to range from $10^{-11}$ to $10^{-8}M_{\odot}{\rm yr}^{-1}$ \citep{1981ApJ...251..139S, 2014A&A...564A..70K}, with the companion mass spanning $10^{1}$ to $10^{2}M_{\odot}$, resulting in a mass-transfer rate range of $\dot{M}_{\rm T}\sim 10^{-16}-10^{-12}M{\odot}{\rm yr}^{-1}$. As discussed in Sec.\ref{sec:acc}, the equilibrium spin period $P_{\rm PA}\propto \dot{M}_{\rm T}^{-3/7}$ and $P_{\rm EP}\propto \dot{M}_{\rm T}^{-1/6}$ are only weakly affected by the mass transfer rate. At most, this allows for the potential emission of FRBs by neutron stars in the propeller stage at low mass transfer rates, as discussed in Sec.\ref{sec:FRB180916}.}

{ 
Note that there are additional effects, caused by accretion, that could prevent the coherent curvature radiation mechanism from operating.  In particular, even if the Alfvén wave reaches the charge-starved region in the presence of accretion,  FRBs may not be generated because the accretion flow disrupts the bunching mechanism. The charged bunches required for the coherent curvature radiation mechanism are formed through the two-stream instability \citep{2023MNRAS.522.4907Y}. However, in the presence of accretion flow, the lifetime  of the charged bunches is significantly reduced, meaning that they may dissipate before an FRB can be generated. While this effect could influence the coherent curvature radiation mechanism, we do not consider it further in this paper, taking it into account would only strengthen our conclusions.}

\begin{figure*}[t]
	\centering
	\subfigure{\includegraphics[width=0.8\linewidth]{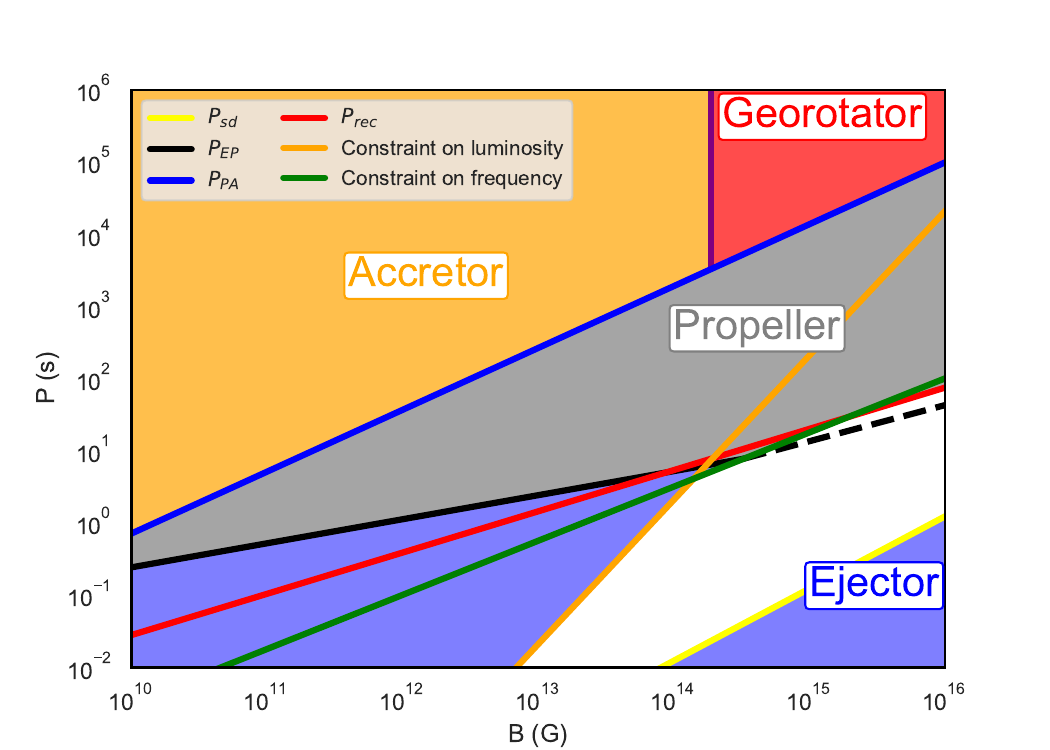}}
	\caption{The neutron star parameters for the magnetic reconnection can be occured and the generation of ${\rm GHz}$ and characteristic luminosity FRBs. Different from decayed Alfvén waves model, the orange line is the condition that magnetic reconnection can occur \eqref{Prec}. { Magnetic reconnection can occur within specific parameter ranges of neutron stars during the propeller phase, and the solid lines have been replaced by the dashed lines in this region to reflect this phenomenon.}}
	\label{condition of reconnection}
\end{figure*}

\subsection{Magnetic reconnection  model}
\par
   In this model, the magnetic reconnection occurs at the light cylinder. Beyond the light cylinder, the current sheet is disrupted by the outwardly propagating low-frequency pulse. Subsequently, the field line tubes with oppositely oriented magnetic fields fall into the magnetic pulse, thereby forming multiple small current sheets. Within each of these small current sheets, a reconnection process takes place through the formation and merging of magnetic islands, giving rise to fast magnetosonic (FMS) waves. These FMS waves are then converted into FRBs, as suggested by \citep{2020ApJ...897....1L}. Hence, this radiation mechanism is dependent on the existence of a current sheet at the light cylinder. Once the current sheet is disrupted, this mechanism ceases to operate. {The condition under which the current sheet can be stabilized is the pressure of magnetic field is balanced by the shock pressure of particles in the current sheet \citep{2020ApJ...897....1L, 2022ApJ...932L..20M}. In the presence of accretion and $R_{\rm m}<R_{\rm L}$ the particles in the current sheet are dominated by accreted matter, in this case the shock pressure of the accreted matter is much larger than the magnetic pressure at the light cylinder $R_{\rm L}$, rather than only at the magnetospheric radius $R_{\rm m}$, where the shock pressure of the accreted matter is balanced by the magnetic pressure, the current sheet cannot be stabilized in the accretion environment.} Therefore, in the binary system, for this radiation mechanism to function, a condition must be met: there should be no contact between the accreted material and the current sheet, which implies $R_{\rm m}>R_{\rm L}$,  Or in other words,  $P<P_{\rm rec}$, where
\begin{eqnarray}\label{Prec}
    P_{\rm rec}&=&\frac{2^{4/7}\pi B^{4/7}R_{*}^{12/7}}{\dot{M}^{2/7}_{\rm T}(GM_{*})^{1/7}c}\nonumber\\
    &\simeq& 5.4 B_{14}^{4/7}\dot{M}_{\rm loss,-8}^{-2/7}M^{4/21}_{\rm c,1} P_{\rm orb, 1}^{8/21}v_{\rm w,8}^{8/7}\,{\rm s}.
\end{eqnarray}
\par
    The luminosity of the FRBs in this model can be estimated as \citep{2020ApJ...897....1L}
\begin{eqnarray}
    L_{\rm FRB}&\sim& \frac{fB_{\rm pulse}B_{\rm L}^{2}R_{\rm L}^{3}}{B_{\rm wind}\tau}\nonumber\\
    &\simeq& 2.1\times 10^{42}f_{-2}\tau_{-3}^{-1}b_{-1}B_{14}^{2}P_{0}^{-1}\,{\rm erg}\,{\rm s}^{-1},
\end{eqnarray}
    where the fraction $f \lesssim 0.01$ of the reconnecting magnetic energy is emitted in the form of FMS waves \citep{2019ApJ...876L...6P,2022ApJ...932L..20M}, $B_{\rm pulse} = bBR_{*}/R$ is the magnetic field of the magnetic pulse \citep{2014MNRAS.442L...9L}, $b\ll 1$ is the dimensionless constant, $B_{\rm wind} = B_{\rm L}R_{\rm L}/R$ is the magnetic field in the neutron star wind, $B_{\rm L}=B_{14}R_{*}^{3}/R_{\rm L}^{3}$ is the magnetic field of neutron star in the light cylinder, and the $\tau\sim 10^{-3}$ is the duration of the observed FRB. 
\par
	For the typical luminosity of a FRB source, $L_{\rm FRB}=10^{42}\,{\rm erg}\,{\rm s}^{-1}$ for example, we can get a upper bound for $P$,
\begin{eqnarray}\label{P_lum_rec}
    P<2.1~f_{-2}b_{-1}B_{14}^{2}L_{\rm FRB, 42}^{-1}\tau_{-3}^{-1}\,{\rm s}.
\end{eqnarray}
It should be noted that the  ``<"  sign in the above equation arises from the condition that $b \ll  0.1$ \citep{2014MNRAS.442L...9L,2020ApJ...897....1L}.
\par
    The characteristic frequency of the emitted waves is determined by the collision time of two merging islands \cite{2020ApJ...897....1L},
\begin{eqnarray}
    \nu'\sim 2\pi c/(\kappa a'),
\end{eqnarray}
    and the size of the islands is $10-100$ times larger than the width of the current sheet $a'$ \citep{2019ApJ...876L...6P}, the prime refers to quantities in the wind frame. The width of current sheet is determined by the neutron star parameters. According to \cite{2020ApJ...897....1L}, the emitted frequency in the observer's frame may be estimated as
\begin{eqnarray}
    \nu & = & 2 \Gamma \frac{\omega^{\prime}}{2 \pi}  =  \frac{1}{\pi \kappa}\left(\frac{r_{e}}{3 \epsilon \zeta c \Gamma}\right)^{1 / 2} \omega_{B}^{3 / 2} \nonumber\\
    &\sim&1.02\times 10^{10}b_{-1}^{5/4}B_{14}^{3/2}P_{0}^{-2}\kappa_{1}^{-1} \zeta_{1}^{-1 / 2} \epsilon_{-1}^{-1 / 2}\,{\rm Hz}
\end{eqnarray}
    where $\kappa\sim 10-100$, $\zeta$ is a few value, $\epsilon\sim 0.1$ is the reconnection rate, and the $\Gamma = (B_{\rm pulse}/4B_{\rm wind})^{1/2}$ is the Lorentz factor of wind, $r_{\rm e}$ is the classical electron radius, and $\omega_{\rm B} = eB_{\rm pulse}/(m_{\rm e}c)$ is is the cyclotron frequency. Therefore, in order to generate ${\rm GHz}$ and characteristic luminosity FRB, the neutron star parameters must be satisfied
\begin{eqnarray}\label{P_GHz_rec}
    P&<&3.2~b_{-1}^{5/8}B_{14}^{3/4}\nu_{9}^{-1/2}\kappa_{1}^{-1/2} \zeta_{1}^{-1/4} \epsilon_{-1}^{-1/4}\,{\rm s}.
\end{eqnarray}
\par
    In summary, the constraints on the parameters of neutron stars are shown in the figure  \ref{condition of reconnection}.  
    Similarly, in the case of this model, a portion of the parameter space within the ejector stage satisfies the constraints, and this space is relatively larger compared to the previous case.The region that meets the constraints is also the area where the magnetic field  is relatively large. This implies that for the Magnetic reconnection model to be viable in the context of FRB generation within the binary system, the neutron star must possess certain specific parameter values within the ejector stage, with a significant magnetic field being a crucial factor. Understanding these parameter ranges is essential for further elucidating the role of magnetic reconnection in FRB production and for better characterizing the possible neutron star scenarios in binary systems that could give rise to these mysterious radio bursts.

    Another type of coherent inverse Compton scattering, as proposed by \cite{2022ApJ...925...53Z}, is described as follows. During a flaring event, crustal oscillations near the neutron star surface may excite low-frequency electromagnetic waves. The X-mode of these waves have the ability to penetrate through the magnetosphere. Bunched relativistic particles, either in the charge starved region within the magnetosphere or in the current sheet outside of it, can upscatter these low-frequency waves, thereby producing GHz emission that powers FRBs. This radiation mechanism encompasses two scenarios.
    In the first scenario, which bears similarities to the coherent curvature radiation described in \cite{2020MNRAS.494.2385K}, the cracking of the neutron star's crust triggers crustal seismic activity. This, in turn, excites Alfvén waves within the magnetosphere. The generation of low-frequency electromagnetic waves follows, with the X-mode waves propagating at a speed approaching that of light. These waves are upscattered at a sufficient altitude where a strong parallel electric field develops, potentially due to charge starvation within the Alfvén wave. Consequently, the conditions necessary for the viability of this radiation mechanism in this scenario are identical to those of the coherent curvature radiation.
    In the second scenario, the emission region is located in the reconnection current sheet region just outside the light cylinder, analogous to the situation described in \cite{2020ApJ...897....1L}. However, the radiation mechanism here operates via coherent inverse Compton scattering by bunches, rather than through the oscillations of colliding magnetic islands within the current sheet. The energy of the particles is related to the magnetic energy released during magnetic reconnection, similar to the magnetic reconnection radiation mechanism. Importantly, if the current sheet is disrupted by accreted matter, this radiation mechanism ceases to exist. Therefore, this scenario is comparable to the magnetic reconnection model.

\section{Case Study}
The long-term periodicity in the bursts of FRB 180916 and FRB 121102 has been observed. This periodicity is believed to likely originate from the orbital period of a binary system, suggesting that the neutron stars driving the FRBs are in binary systems and likely contains massive companions. Additionally, the RM inversion evolution of FRB 20201124A and FRB 20190520B  can also be well explained within the context of binary systems. However,  when the neutron stars generating the FRBs are undergoing accretion, the radiation process of FRBs within the magnetosphere is bound to be affected by the accretion flow. What conditions must be met for the radiation process to proceed normally? In this section, we will apply the general theory from the previous section to carefully examine this issue based on these special FRBs.

\subsection{FRB 20201124A}
\cite{2022NatCo..13.4382W}  proposes that repeating FRB 20201124A originates from a neutron star/Be star binary system. In this model, the neutron star orbits a Be star with a decretion disk. When approaching the periastron, the interaction between radio bursts and the disk causes observed phenomena like RM variation. The model can reproduce the RM evolution of FRB 20201124A and explain that of FRB 20190520B. 

\begin{figure*}[t]
	\centering
	\subfigure{\includegraphics[width=0.45\linewidth]{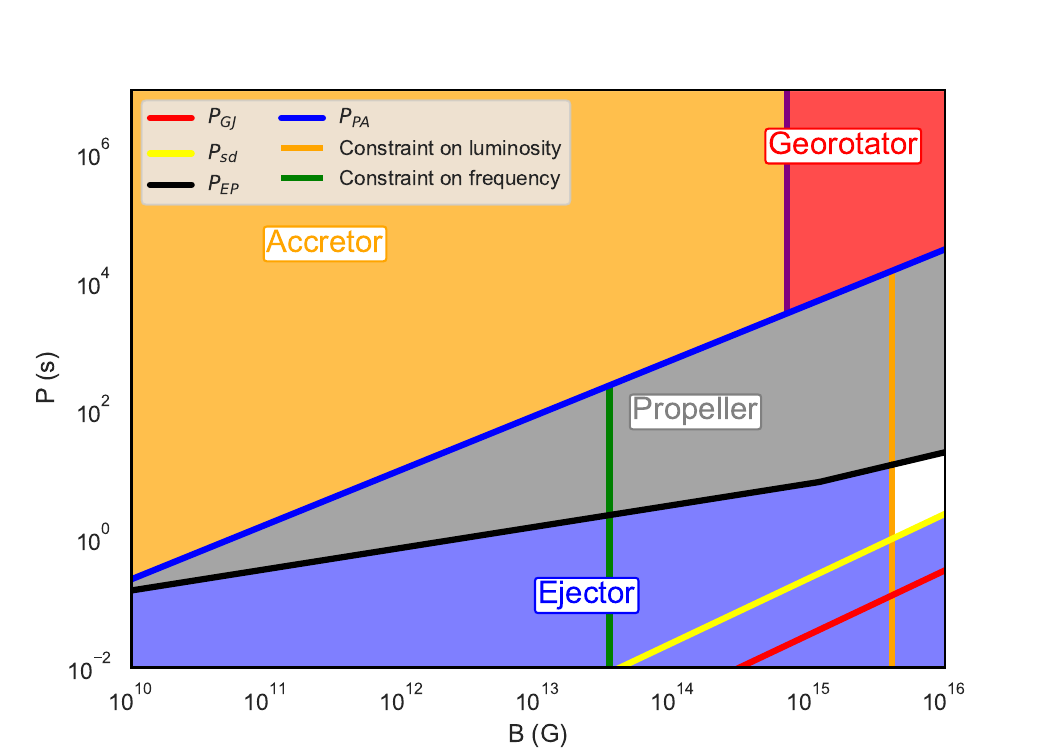}}
	\subfigure{\includegraphics[width=0.45\linewidth]{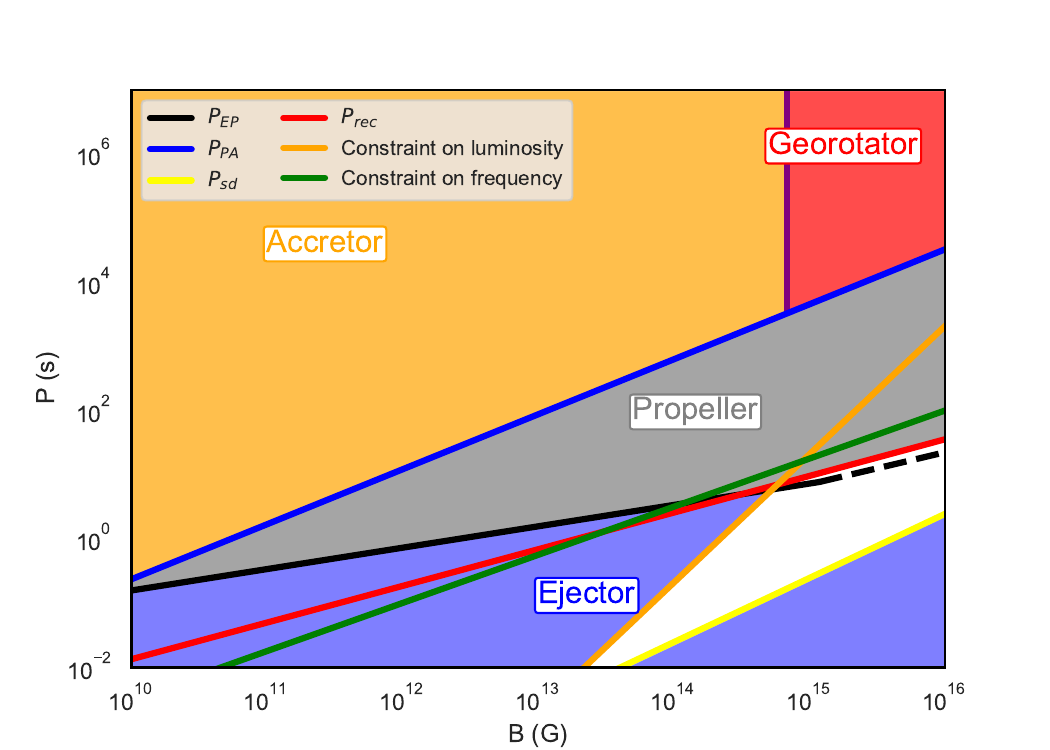}}
	\caption{The neutron star parameters for the coherent curvature radiation mechanism and the  magnetic reconnection for the model of \cite{2022NatCo..13.4382W}. Here,  for FRB 20201124A, the luminosity can reach $10^{43}\,{\rm erg}\,{\rm s}^{-1}$, and the active timescale is more than 4 years.}
	\label{model4disk}
\end{figure*}

In the model proposed by \cite{2022NatCo..13.4382W}, the companion star has a mass of $M_{\rm c} \sim 8M_{\odot}$, a mass loss rate of $\dot{M}_{\rm loss} \sim 10^{-10}M_{\odot}{\rm yr}^{-1}$, and an orbital period of $P_{\rm orb} \sim 80\,{\rm d}$. The companion star is a Be star surrounded by an accretion disk that extends from the stellar surface to larger radii. The density of the disk at the stellar surface is $\rho_0=3 \times 10^{-14}$ g cm$^{-3}$, and it decreases with radius as $r^{-4}$. Using these parameters, we analyzed the parameter space required for a neutron star to produce FRBs. The results, presented in Fig.\ref{model4disk}, show that the coherent curvature radiation mechanism is only feasible within a very narrow parameter space within the ejector stage.
 However, the magnetic reconnection mechanism operates with a significantly larger parameter space of the ejector stage
and it is even viable within a small portion of the propeller stage's parameter space.  In any case, to produce FRBs without being significantly affected by the accretion process, a magnetic field strength of $B > 10^{15}~{\rm G}$ is required for a  neutron star with spin period on the order of one second. Even for a rapidly spinning neutron star with a period of ten milliseconds, the magnetic field strength must still  $>10^{13}~{\rm G}$.  Anyway, the neutron star  needs to have a strong enough magnetic field, that is, it must be a magnetar.

\subsection{ FRB 20180916B}\label{sec:FRB180916}
   In the study of \cite{2024ApJ...967L..44L}, a binary model was developed for  the physical origin of the periodic activity of FRB 20180916B. This model assumes a massive star binary system with an orbital period of about $P_{\rm orb}\sim 3000\,{\rm d}$. The binary contains a slowly rotating neutron star with a rotation period of around $P\sim 16.35\,{\rm d}$ and a magnetic field strength of about $B\sim 10^{15}\,{\rm G}$, along with a massive star of approximately $M_{\rm c}\sim 30M_{\odot}$ and a significant mass loss rate of $M_{\rm loss}\sim 10^{-9} M_{\odot}\,{\rm yr}^{-1}$. This configuration is proposed to explain the many observational features of FRB 20180916B. Notably, the model can account for the complex variations and reversals in the Faraday rotation measures. The periodicity of the FRB is thought to come from the long rotation of the neutron star, and the change in the rotation measure is due to the mass loss of the massive star in the binary system. While this model provides a useful framework for understanding FRB 20180916B, 
   it relies on several assumptions and conditions that warrant further scrutiny, particularly regarding the role of accretion and its potential influence on the feasibility of the model.
   In the following analysis, we will evaluate this model in light of the constraints established in the previous section. This will help assess the model's potential to offer a more comprehensive understanding of the mechanisms underlying FRBs.
   \begin{figure*}[t]
   	\centering
   	\subfigure{\includegraphics[width=0.45\linewidth]{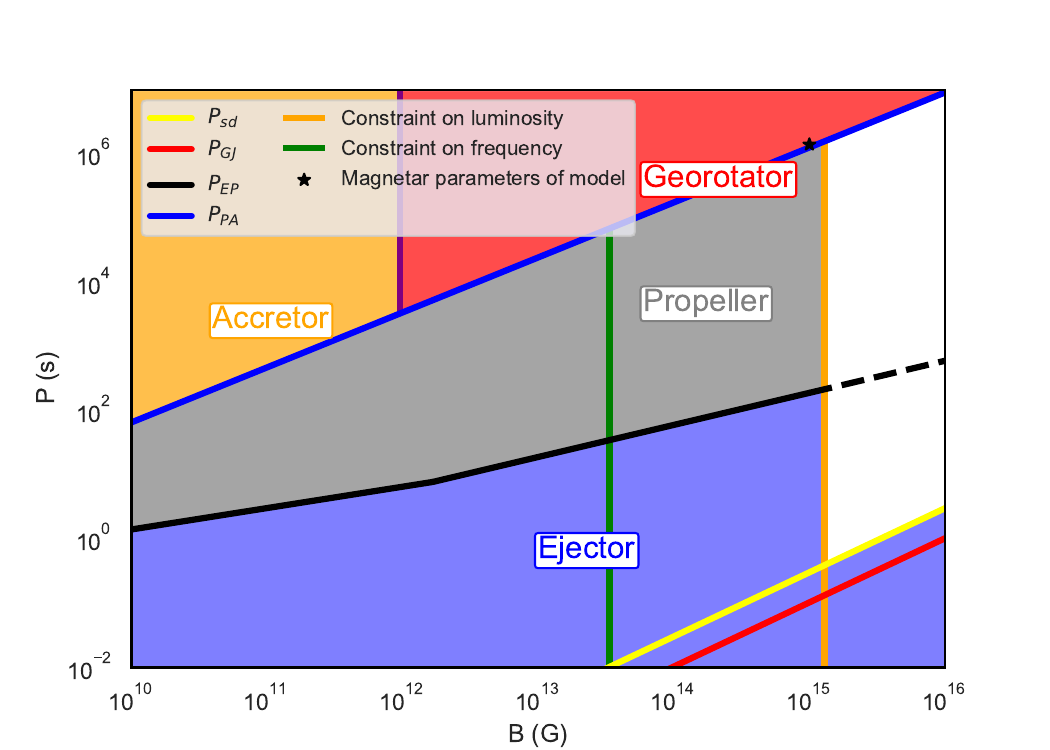}}
   	\subfigure{\includegraphics[width=0.45\linewidth]{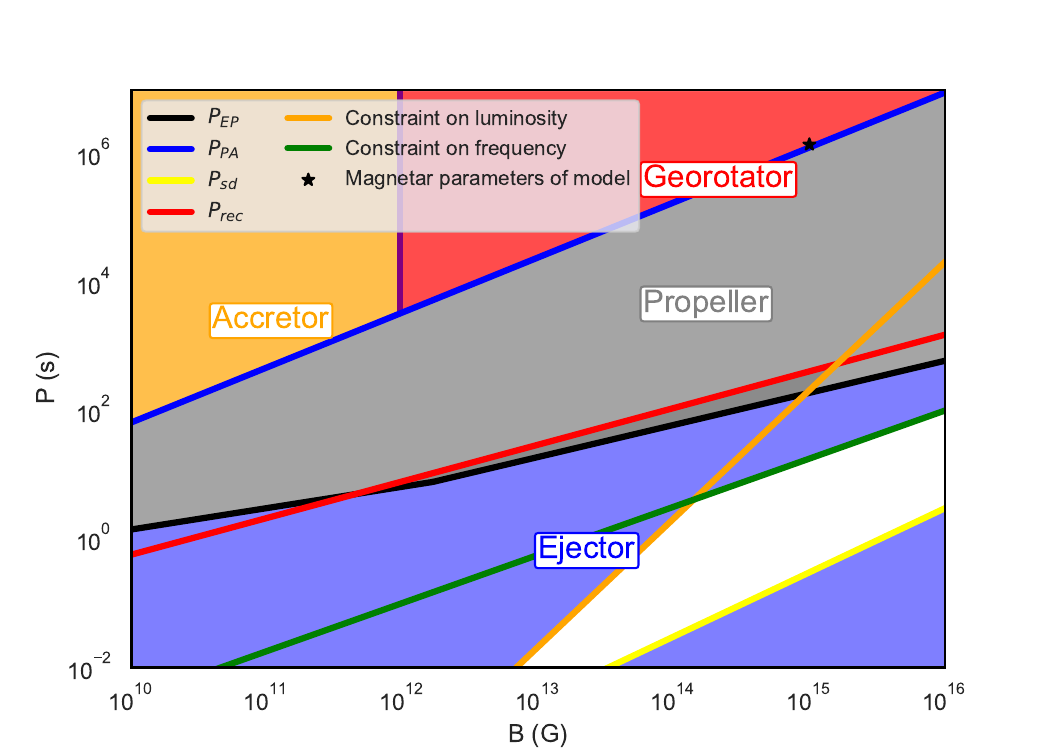}}
   	\caption{The neutron star parameters for the coherent curvature radiation mechanism and the  magnetic reconnection for the model of \cite{2024ApJ...967L..44L}. Here,  for FRB 20180916B, the luminosity can reach $10^{42}\,{\rm erg}\,{\rm s}^{-1}$, and the active timescale is more than 6 years.}
   	\label{model1}
   \end{figure*}
    In this case, we calculated the characteristic radii using equation \eqref{radii} and found that $R_{\rm co} > R_{\rm m} > R_{\rm G}$, indicating that the neutron star accretion is in the georotator stage. At this stage, the particle density, as derived from equation \eqref{n_G}, is significantly higher than the critical density given by equation \eqref{eq:ncR}. Therefore, if the radiation mechanism at play is the coherent curvature radiation, the production of FRBs is not feasible under these conditions.
    However, if the neutron star's magnetic field well exceeds $10^{15}~{\rm G}$, the accretion state transitions into the propeller stage, where the density of the accreted matter can drop below the critical density $n_{\rm c}$ required for coherent curvature radiation. On the other hand, the magnetic reconnection mechanism becomes unviable in this scenario because the magnetospheric radius $R_{\rm m}$ is smaller than the light-cylinder radius $R_{\rm L}$, causing the accreted matter to disrupt the structure of the current sheet, which is essential for this mechanism.
    Therefore, within this model, if the coherent curvature radiation mechanism is responsible for FRB generation, a magnetic field strength  well exceeds $ 10^{15}~{\rm G}$ is required for the neutron sta, to produce FRBs. This condition ensures that the density of accreted matter remains below the critical threshold, allowing coherent curvature radiation to occur. The results of this analysis are  illustrated in Fig.\ref{model1}. One sees that the neutron star needs to be a slowly rotating magnetar with an extremely strong magnetic field. 

We have noticed that, at earlier time, \cite{2021ApJ...918L...5L} have proposed that a Be/X-ray binary system as the source of the periodic repeating FRB 20180916B. When an NS in the system accretes material from the Be star disk, it causes spin evolution. Starquakes occur when the crust stress reaches the critical value, producing FRBs  through coherent curvature radiation in the magnetosphere. The interval between starquakes varies depending on the NS's position relative to the disk. The free-free absorption of the Be star disk leads to a frequency-dependent active window for FRBs. This model can account for the observational features of FRB 180916B, including its activity window and dispersion measure contribution. However, through the above analysis, it can be seen that when a neutron star is in the accretor stage and accreting the disk of Be companion, there are no suitable conditions at all to generate FRB radiation within the magnetosphere.

\section{Summary and discussion}
In this study, we explored the impact of accretion on the radiation mechanisms responsible for generating fast radio bursts (FRBs) in binary systems. The presence of accretion flows from a massive companion star introduces additional plasma into the magnetosphere of a neutron star, which can significantly alter the conditions necessary for coherent radiation mechanisms. We specifically analyzed the feasibility of two popular FRB emission mechanisms within this context: the coherent curvature radiation mechanism and the magnetic reconnection mechanism.

Our analysis shows that the coherent curvature radiation mechanism is only viable within a very narrow parameter space during the ejector stage of the neutron star's evolution. In this stage, the density of the accreted plasma remains below the critical threshold required for the formation of a charge-starved region, allowing the Alfvén wave-induced coherent curvature radiation to proceed. However, even within this favorable stage, the required magnetic field strength must well exceed \(10^{15}\,\mathrm{G}\) for a neutron star with a rotation period of around one second to produce observable FRBs. For faster-spinning neutron stars with periods on the order of tens of milliseconds, a lower magnetic field of  \(\gtrsim10^{13}\,\mathrm{G}\) may suffice. Outside of this narrow ejector stage, the dense accreted plasma disrupts the conditions necessary for this radiation mechanism, rendering it ineffective.

In contrast, the magnetic reconnection mechanism exhibits a broader viable parameter space. Our calculations indicate that this mechanism remains feasible across most of the ejector stage and even extends into a portion of the propeller stage. However, this mechanism ceases to operate if the magnetospheric radius falls below the light cylinder radius, as the accreted material would then interact with the current sheet, preventing efficient magnetic reconnection. To sustain the conditions necessary for this mechanism, the neutron star must maintain a sufficiently high magnetic field strength and spin rate to prevent the disruption of the current sheet by accreted matter.

Applying our findings to recent binary models proposed for specific repeating FRBs, such as FRB 20180916B and FRB 20201124A, we find that the radiation mechanisms must account for the presence of accreted matter from the companion star. In the case of FRB 20201124A, a Be star companion with a decretion disk significantly affects the plasma environment of the neutron star. Our analysis suggests that coherent curvature radiation is only feasible within a limited portion of the ejector stage, whereas the magnetic reconnection mechanism can potentially operate in both the ejector and early propeller stages. The magnetic field strength of the neutron star must be sufficiently strong to withstand the influence of the accreted material and maintain the conditions necessary for FRB generation.

In summary, our work highlights the stringent constraints on neutron star parameters required to produce observable FRBs in binary systems. Both radiation mechanisms considered here demand a magnetar-level magnetic field strength to overcome the impact of accretion and maintain viable emission conditions. These findings suggest that neutron stars capable of generating FRBs within binary systems are likely to be magnetars with exceptionally strong magnetic fields. Future observations, particularly of FRBs with periodic activity and evolving rotation measures, will be crucial for refining the parameter space of neutron stars in binary systems and improving our understanding of the underlying FRB generation mechanisms.
{Furthermore, our results  indicate that the optimal stage for generating FRBs is in the ejector stage. 
At this stage, the interaction between the particle winds from the magnetar and its companion star can generate shock waves, which are known as bow shocks \citep{1971SPhD...15..791B,2002ApJ...575..407C}. Magnetospheric activity, such as magnetic reconnection or starquakes, can launch MHD shocks which arise from the nonlinear steepening of compressive waves (e.g., Alfvén waves or fast magnetosonic waves \citep{2003MNRAS.339..765L}).   It can be expected that these shock waves will accelerate electrons to produce synchrotron radiation, which is a possible FRB-associated signal worth people's attention.}

\section*{Acknowledgments}
This work is supported by National Key R\&D Program of China (grant No. 2023YFE0117200).
Can-Min Deng is partially supported by the National Natural Science Foundation of China (grant No. 12203013) and the Guangxi Science Foundation (grant Nos. AD22035171 and 2023GXNSFBA026030).

\bibliography{ref}

\end{document}